# Path to Low Cost Microfluidics

*Alexander Govyadinov, Erik Torniainen, Pavel Kornilovitch and David Markel*
**Hewlett-Packard, Corvallis, Oregon, USA**

## Abstract

*The paper describes a novel concept for a low cost microfluidic platform utilizing materials and processes used in low cost thermal inkjet printing. The concept re-purposes the jetting elements to create pumps, mixers, and valves all necessary components for the transport of fluids in a broad range of microfluidic applications.*

## Introduction

There is a great need in reliable and scalable microfluidic technologies that can serve diverse engineering, biomedical, and environmental applications. Microfluidic hardware includes medical research systems, point-of-care and molecular diagnostic systems, high pressure liquid chromatography chips (HPLC), digital polymerase chain reaction (PCR) chips, micro-needles and other devices. This highly fragmented market is estimated at $7.2B today but predicted to grow at 20% annually to $21B in 2020, even in the absence of a dominant killer application [1][2]. Wider adoption of these technologies requires compact inexpensive systems that seamlessly integrate diverse microfluidic, sensing, software and communication components.

One path to overcoming this technical challenge is via expansion of inkjet hardware platforms, for example the one that has been powering Hewlett-Packard's thermal inkjet printing (TIJ) business for the last 30 years [3]. Economies of scale and high level of fluidic and electrical integration achieved in TIJ printheads promise low cost microfluidics when it is repurposed for other applications.

At the heart of HP TIJ technology is the TIJ resistor. When heated by microsecond pulses, the resistor boils the surrounding fluid that expands and ejects a fluid droplet into ambient. More generally, the TIJ resistor can be viewed as a power source that can deliver localized pressure bursts at any point of a microfluidic system. It was recently demonstrated that the TIJ resistor can also work as a pump in close microchannels, pumping fluids with viscosity as high as 16 cP in channels as narrow as 5 µm [5]. Beyond pumping, the TIJ resistor can perform other functions such as valving, mixing, and heating.

The aim of this article is to demonstrate that the HP TIJ technology can provide many of the functional building blocks necessary to enable low-cost microfluidic systems.

## Methods

The potential for Thermal Inkjet technology to consolidate multiple microfluidic devices in one platform using CMOS and microfluidic MEMS fabrication processes is described elsewhere [3]. We have used Computational 3D Fluid Dynamic CFD3 code [4][5] and a 1D model for numerical simulation [6] of the microfluidic devices for evaluation of their performance. A massless particle seeding approach has been used for flow visualization, particularly for the calculation of mixing efficiency of micro-mixers.

TIJ-based microfluidic system typically includes microfabricated SU8 microchannels 5-25 µm wide and 11-22 µm tall with a linear density 600-1200 resistors per inch (see Figure 1). Every channel typically includes a microheater (a thermal inkjet resistor) which is 2-20 µm wide and 12-45 µm long located near one or other end of the channel and is activated using traditional TIJ fire pulses [5].

Fabricated devices were filled with different fluids ranging from water to biological liquids such as whole blood. The fluid's viscosity varied from 0.8 to 16 cP at operation temperature [5].

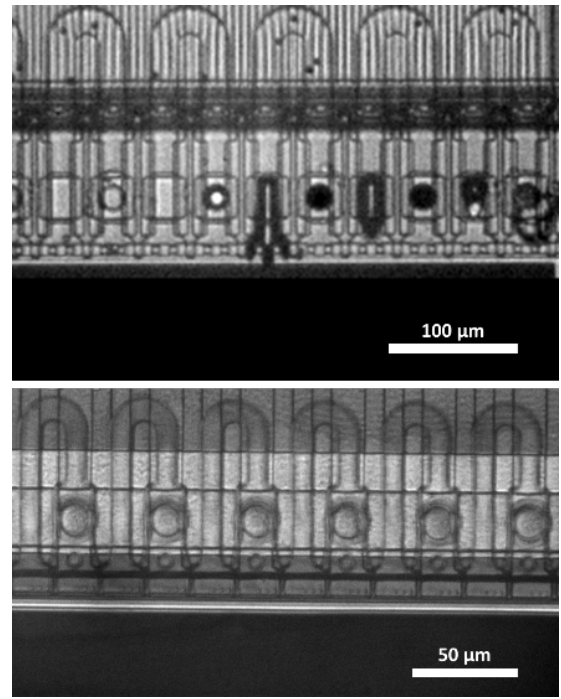

**Figure 1:** *Shows typical devices with 600 (upper) and 1200 (lower) channels per inch. The 600 channel per inch device contains Ø 4.0 µm tracers. Black area shows fluid reservoirs and drive bubbles are shown with dark contours*

The fluid flow was observed using conventional optical microscopy with coaxial illumination. The optical system enabled 5x-250x magnification where 50x magnification was used for most of characterizations. We used two methods for the microfluidic system video characterization: stroboscopic imaging and high speed imaging. For stroboscopic imaging a high intensity stroboscopic LED illumination source with pulse duration from 200 ns to 2.5 µs has been used. The stroboscopic video was recorded with a high sensitivity µEye USB camera. High speed video-recording was used for real time particle tracking using high intensity Xe-lamp and Vision Research high speed camera Phantom v1610 with frame rates between 25,000 and 1,000,000 frames per second. Fluid flows were visualized by 1.6-4.0 μm diameter polystyrene spheres added to the fluid. Particle images were used to calculate their displacement and characterize instantaneous and average flow velocities.

## Bubble-Driven Inertial Micropump

Moving fluids in micro-systems is the most important function in microfluidic systems and this requires pumps that can accurately and reliably distribute fluid in the presence of particles and air bubbles. Pumps must deliver flow to fluidic components such as mixers, filters, reaction chambers, resistors, etc. through micro-channels in potentially complicated geometries over a wide range of fluid properties. This pumping capability is the foundation of many microfluidic systems such as inkjet, chemical micro-reactors, and lab-on-a-chip devices. Previously, many types of pumps have been developed to perform pumping in micro-systems [7] and pump technologies continue to be developed.

The challenging problem of microscale pumping has been approached in a number of ways. A wide variety of micropumps have been used including active and passive pumping schemes [7-12]. While these pumps are effective for pumping in specific circumstances, general disadvantages of these pumps are moving parts, complicated geometries, large sizes, and difficulty in fabrication or constraints on the fluid properties.

The idea of an inertial pump was based on using Thermal InkJet (TIJ) resistors to circulate the fluid around TIJ microchambers. Microcirculation would reduce the undesirable consequences of water evaporation, a problem linked to increased viscosity and clustering of pigment. It was conjectured that a standard TIJ resistor placed asymmetrically within a microfluidic channel without a nozzle opening might induce a net fluid flow by repeated generation and collapse of thermal vapor bubbles. Such a system was built and indeed a net fluid flow in the channel was observed, as shown in Figure 2. It quickly became clear that a TIJ resistor in an asymmetric channel constitutes a completely novel type of integrated micropump. With no moving parts and fabricated with standard fab processes, complex integrated devices could now be built in hundreds on a single chip forming a type of micro distribution system that expanded beyond inkjet.

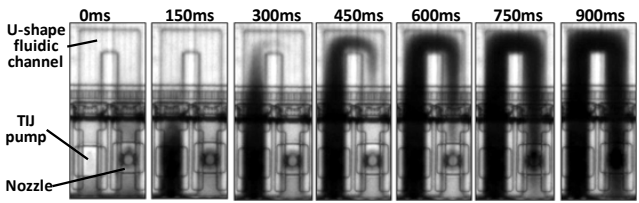

**Figure 2:** *Time series images of black pigment particles being pumped by the enclosed in microfluidic channel resistor marked "TIJ pump", which generates a net flow around a U-shaped microfluidic channel.*

### *Physical principles beyond pump operation*

The operating principle behind inertial pump can be understood on a simplified one-dimensional model illustrated in figure 3(a). The pump contains a microheater at coordinate $x_0$ inside a microchannel which separates two reservoirs at an ambient pressure $p_0$. The microheater creates a high pressure region (a model vapor bubble) of pressure $p_{vr} > p_0$, which induces outward flow in the long and short arms of the microchannel (the stage of bubble expansion). The short arm contains less fluid, so it accelerates more rapidly into the left reservoir, while the long arm has more fluid and it accelerates more slowly into the right reservoir, Figure 3(c). Very rapidly, the pressure in the bubble drops below $p_0$ and further expansion proceeds mostly by inertia against a negative pressure difference, Figure 3(d). Both fluid columns lose their mass into reservoirs, dissipating part of mechanical momentum. Crucially, the shorter arm loses its mass faster, which makes it easier to stop by external pressure, while the longer arm has more inertia and takes longer to stop. (Hence the term "inertial" pump.) When outward flow stops and is replaced by inward flow, the short arm reverses direction earlier while the long arm continues outflow a bit longer, Figure 3(e). Eventually, the two flows collide to the right of the microheater location, which amounts to a net flow and constitutes the primary pumping effect, Figure 3(f). Additionally, since the short arm has more time to accelerate under external pressure, it will have a larger mechanical momentum at collision. As a result, there is postcollapse flow from left to right, which constitutes the secondary pumping effect, Figure 3(g).

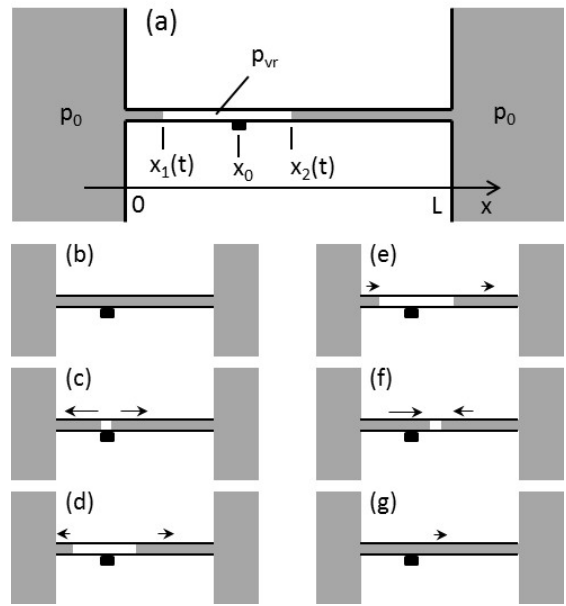

**Figure 3.** *(a) Schematic geometry of an inertial micropump. (b)–(g) Phases of the expansion-collapse cycle. (b) In the starting state, the fluid is at rest. (c) The microheater (black square) creates a high pressure vapor bubble. A positive pressure difference pushes the fluid out of the channel. (d) The vapor bubble pressure quickly drops below atmospheric and the fluid decelerates under a negative pressure difference while moving by inertia. (e) The short arm turns around, while the long arm is still moving out of the channel. (f) The two fluidic columns collide at a point shifted from the starting point of the expansion: the primary pumping effect. (g) Since the shorter arm has a larger momentum at collision, the total postcollapse momentum is nonzero: the secondary pumping effect.*

In the one-dimensional approximation, a pump state is completely described by the locations of vapor-fluid interfaces $x_{1,2}(t)$. A dynamical model for $x_{1,2}(t)$ that details the effects described above was developed and studied in [6].

### *Computational Fluid Dynamics simulations*

While the semi-analytical model provides an understanding of the physical basics for the pumping mechanism, it is too simplistic for the three-dimensional geometries of real microfluidic devices.

We have therefore performed extensive numerical simulations using Computational Fluid Dynamics (CFD) tools developed in the Modeling and Analysis group (EMAG) of HP Corvallis [4],[5] and [12]. The simulations have been used to predict operation of the pump and guide our design. After many years of droplet ejection simulations, the modeling group has accumulated extensive experience of inkjet boiling and piezo actuation. We have applied these models to study the factors involved in the inertial pumping event.

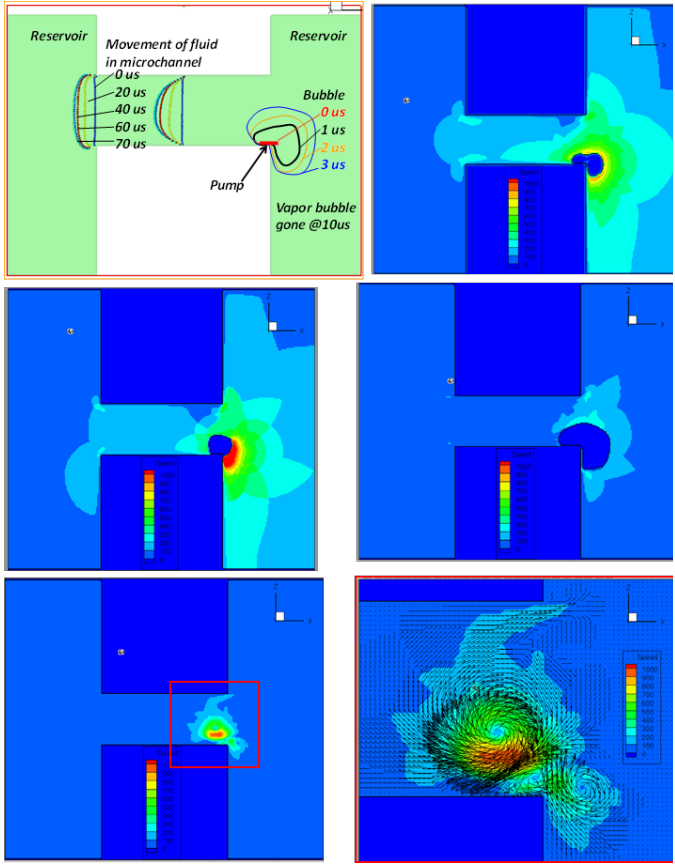

**Figure 4:** *CFD simulation of inertial pump with a small vapor bubble actuator located on the right side of a microchannel. Temporal motion of vapor bubble and fluid particles is shown in the top left figure. The other five panels display velocity fields during an inertial pumping event. Colors represent speeds from 0 to 1000 cm/s at 1 µs (top right), 3 µs (top right), 7 µs (middle left), and 10 µs (bottom left). A close-up view of the velocity vectors at 10 µs is shown in the bottom right. (The close-up area is indicated by the box in the bottom middle image.)*

Figure 4 shows a typical CFD model of a straight microchannel with a thermal inkjet functioning as the inertial pump actuator. This simple geometry is the basic design that we have used to study the inertial pump, because geometrical complications are minimized in this arrangement. The vapor bubble, located on the right side of the channel, rapidly expands and then collapses to create a pumping motion from right to left in the figure. Note that the resultant motion of the particles in the microchannel is delayed by many microseconds after the collapse of the vapor bubble.

Using the CFD simulations we can gain insight into the fluid details such as pressure and velocity fields. In Figure 4, we show the instantaneous velocity fields in the microchannel during a pumping event. First, there is rapid expansion of the drive bubble between 0 and 3 microseconds followed by a collapse event which concludes at roughly 8 µs. Note that the expansion and collapse times may shift depending on geometry, fluid properties, and bubble details, but the expansion phase is typically shorter than the collapse phase. The importance of the temporal asymmetry of the pulse actuation will become clear when we look at piezo-actuation below.

Figure 5 shows that there is high velocity outward from the vapor bubble during the initial expansion followed by relatively low velocity at the maximum bubble extent and very high velocities at the final collapse. CFD simulations show that when the bubble collapses there is a high speed jetting event which forms vortices. These vortices can survive for a number of microseconds as they move down the channel in the direction of net flow (toward the long arm of the channel) before they are dissipated by viscosity. Note that these vortices also provide effective mixing in the vicinity of the bubble collapse, while fluid further away moves laterally as we would expect from pressure driven flow (Poiseuille flow). This mixing effect is shown in Figure 5 where yellow particles are mixed in the vicinity of the pumping resistor and blue/green particles downstream from the vapor bubble move laterally.

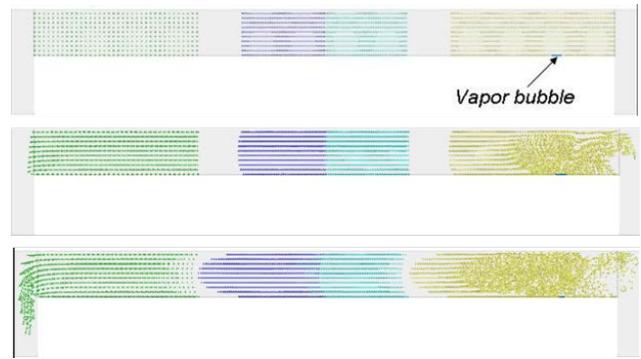

**Figure 5:** *Particle motion in a straight channel during multiple pulsing by an inertial pump located at the right side of the microchannel (blue line). Reservoirs are located to the right and left sides of the channel. The top image is at time zero, the middle image is at 40 µs (after first bubble firing) and the bottom image is after 6 firings at 200 µs.*

We have also looked at the effects of resistor location and vapor bubble strength (see Figure 6). As predicted by the semi-analytical model, the fluid is pumped in the direction of the long column of fluid. If the resistor is placed in the middle of the channel then there is no net pumping effect. If the drive bubble is placed too close to the reservoir, the pumping effect is reduced because the vapor bubble spills into the reservoir rather than into the microchannel. As we increase the size of the drive bubble, the amount of pumping increases as expected. Because of the inertial nature of the device, the operation speed, average flow rate and pressure will be limited by the time constant of the fluidic circuit. Typically, small/micro-scale fluidic circuits with small liquid masses have higher operation rates. We demonstrated operation speeds of 1 Hz-10 kHz with 0.1-10 pL per pulse net flow in typical for inkjet microfluidic channels.

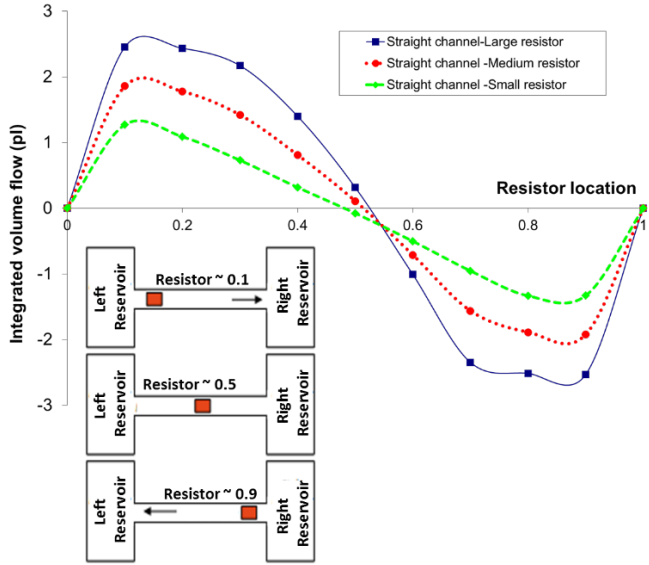

**Figure 6:** *Amount of fluid pumped in a straight channel as a function of resistor location and resistor strength. Resistor location 0 is a resistor located near the left reservoir and resistor location 1 is a resistor location near the right reservoir. The small images in the bottom left indicate resistor locations as well as direction of net flow.*

Inertial pulse dynamics is very fast. The bubble lifetime is about 10 us, and the entire cycle, including post-collapse flow completes within less 100 us (for typical TIJ channel sizes). Recent advances in high-speed imaging technology made it possible to image single pulse dynamics with microsecond resolution and measure displaced volume as a function of time using tracer beads. Representative results are shown in Figure 7.

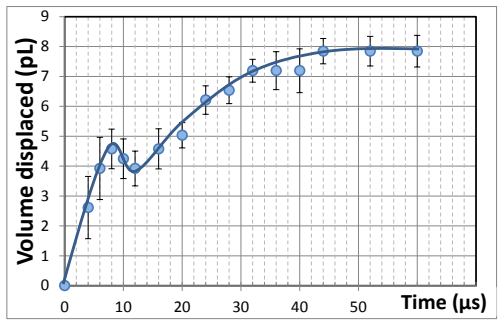

**Figure 7:** *The displaced volume for the case shown in Figure 1 upper as a function of time. The dip at 10µs is due to the drive bubble collapse and is predicted by simulation [5,6].*

### *Micropump piezo-actuation model*

Thermally generated vapor bubbles are not the only possible driving mechanism of the inertial pump. Another possible driving method is mechanical compression by a piezo element, external impact or other actuator types. Because of practical importance, we have run a number of piezo-actuated simulations as shown in Figure 8 and Figure 9.

Figure 8 shows a U-shaped geometry that was used to study piezo-actuation. In this geometry, the piston moves from the chamber floor at the start of a cycle to near the top of the chamber (maximum displacement) at half-cycle time and returns to the chamber floor at the end of the cycle. This piezo-motion results is pumping similarly to actuation by vapor bubbles.

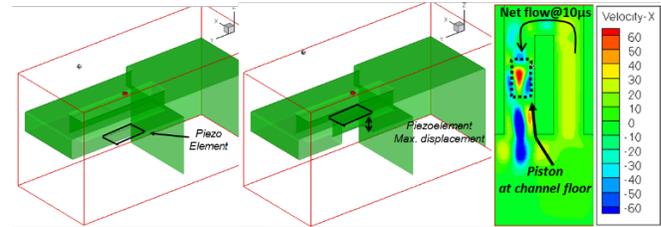

**Figure 8.** *U-shaped recirculation chamber with a piezo-actuated inertial pump. Geometry is shown in the left image. The initial location of the piezo-actuator s on the floor of the chamber (left image), and maximum displacement of the piezo-actuator is near the top of the chamber (middle image). The final piezo location is the same as the initial position. Right image shows the velocity field at 10 µs after the piezo-actuator has completed a cycle.*

In piezo-actuation both phases of the pumping cycle, the expansion and the collapse, can be controlled electrically. Thus piezo-actuation provides more flexibility for pump design and operation than thermal actuation. Figure 9 shows a number of piezo-pumping scenarios that have been investigated.

Analogous to pumping with vapor bubbles, if the piezo element is located in the middle of the geometry then no net recirculation is created. Because we can control the piezo displacement, we see that this effect is true for symmetric or asymmetric drive pulses.

Moving the piezo element away from a symmetric location creates net flow. Through numerical simulations, we have seen that it is necessary to use asymmetric drive pulses at asymmetric locations to achieve pumping effect. Also, pumping direction can be reversed by changing the drive actuation from a high-velocity upstroke with low-velocity downstroke [ $t_1 < ( t_2 - t_1 )$, see the green line in the bottom image of Figure 9] to low-velocity upstroke with high-velocity downstroke [ $t_1 > ( t_2 - t_1 )$, see the blue line in the bottom image of Figure 9].

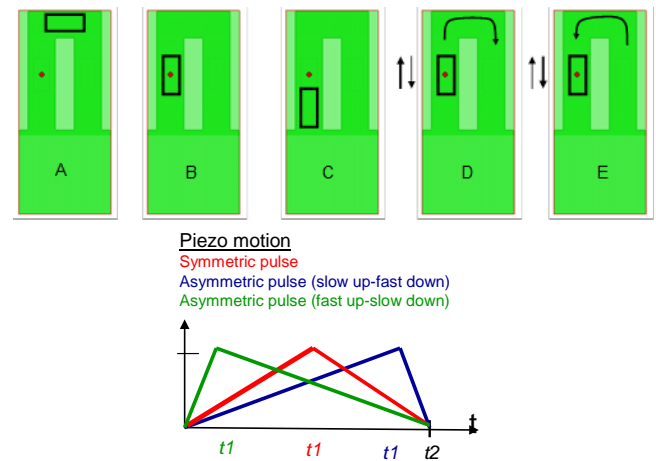

**Figure 9:** *Piezo simulations of symmetric geometries (A) and asymmetric geometries (B-E) using different drive pulses shown at right. The box in the image indicates the piezo element and the arrows in the channel show the net pumping effect. Geometry B is a modestly asymmetric design, geometry C is an aggressively asymmetric design. Geometry D shows the pumping effect for*

*the fast upstroke-slow downstroke piezo pulse (green pulse in bottom graph) and geometry E shows the pumping effect for the slow upstroke-fast downstroke (blue pulse in bottom graph).*

Fluid inertia becomes more prominent at macroscale (large Reynolds numbers). If pumping is indeed due to asymmetry in inertial properties, then it should be possible to demonstrate the effect on macroscopic physical devices. One such device was created and is shown in Figure 10. It consisted of a plastic tube connected to a water reservoir by both ends. The tube was periodically hit with a hammer, and the mechanical impact squeezed the tube and pushed the fluid inside in two opposite directions (expansion phase).

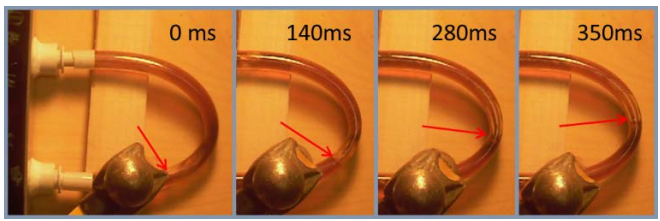

**Figure 10:** *Macro-scale physical device that demonstrates mechanical pump moving water in ¼ inch diameter tubing using a hammer. Repeated hits close to one end of the tube generate quasi-stationary flow in the counter-clock-wise direction, which is visualized by a moving bubble (marked by arrows).*

The collapse phase was driven by elastic rebound of the tube. If the tube was hit closer to one end, quasi-stationary flow from the shorter arm toward the long arm was observed, in accordance with the inertial mechanism. When the tube was hit at the center with the same intensity no net flow resulted, as expected.

The experimental tests have validated the semi-analytical and CFD modeling and demonstrated that the inertial pump is operational and can be used for recirculation and mixing of fluids as it shown in Figure 2. With the standard size of the resistor of 15x30 µm in a fluidic chamber of 22x17 µm cross-section, we achieved a net flow of 0.4-10 pL/pulse or 1-25 µm/pulse displacement. This flow rate is a function of liquid properties, the firing energy and firing frequency of the pump resistor [5].

## Other components

### *Mixer*

A crucial component of an integrated microfluidic system is the micromixer. Chemical, biological, and medical application experts express frustration at the lack of understanding, scalability, and flexibility of current micromixers (see for example [13]). For mixing, current technologies generally require complicated geometries to increase the effectiveness of diffusive mixing. The geometric complexity of these designs increases pressure losses in the system and makes fabrication difficult [14]. CFD simulations show (see Figure 5) that the same TIJ resistors can be used for both pumping and mixing. This mixing method requires no special geometries and produces mixing rates many times that of simple diffusion.

Figure 11 and Figure 12 illustrate our fundamental understanding of mixing on microscale. We are able to model various mixing conditions along with a relevant mixing metric to describe the efficiency of micromixing and its fundamental mechanisms [15].

An active microscale mixer that utilizes the inertial pump and an alternating firing pattern for two side-by-side resistors in a fluidic channel (Figure 11) to drastically improve mixing efficiency when compared to current mixing techniques.

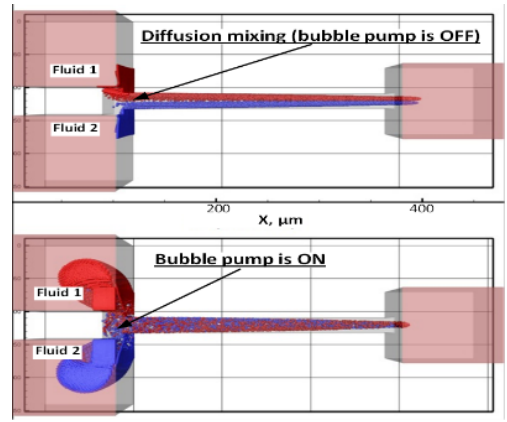

**Figure 11:** *Mixing comparison between pressure driven flow (top) and active mixer (bottom).*

A particle seeding feature applied to CFD3 model (Figure 12) enabled quantitative characterization of mixing by statistical analysis of massless particle in mixing zones which was used for comparative analysis of developed mixers with other mixing technologies. Using particles allowed us to track mixing without the effects of diffusion which is useful because diffusion can artificially boost micro-mixing.

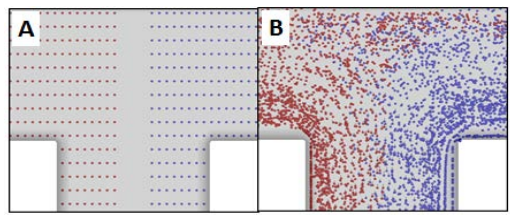

**Figure 12**: *Model of a pressure driven microfluidic system Fig. 12 with a flow rate of 0.1 m/s depicting particle tracking used for mixing factor calculation: (A) mixing zone at time 0, (B) mixing zone at 500µs.*

### *Dynamic valve*

A popular valving technology in microfluidics is commercialized by Fluidigm pneumatic polymeric (PDMS) valve developed by Quake's group [10],[16]. This valve is expensive, does not work with rigid microfluidics and not easily integrated onto chips because of the complexity of an external pneumatic drive. Also, PDMS has material compatibility issues because it absorbs organic solvents. We are in the early stages of developing a flow valve that uses pumps to control the direction of flow as shown in Figure 13 and Figure 14. The goal of the valve is to have no flow into one channel while continuing to pump into another channel. There are three major parameters we use to control flows produced by every pump: resistor size and location and operation frequency. In practice, all of these parameters are beneficial for dynamic valve implementation.

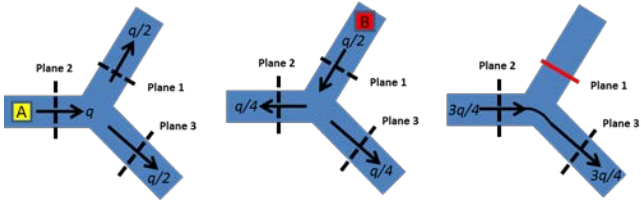

**Figure 13:** *Principle of operation for valve in a Y-channel. First resistor 'A' is fired, then resistor 'B' is fired pushing all of the fluid out of the branch of the Y-channel. The net flow is show in the graph in Fig 14 on the bottom.*

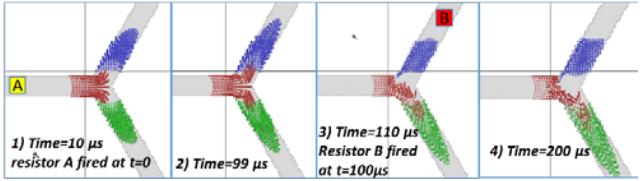

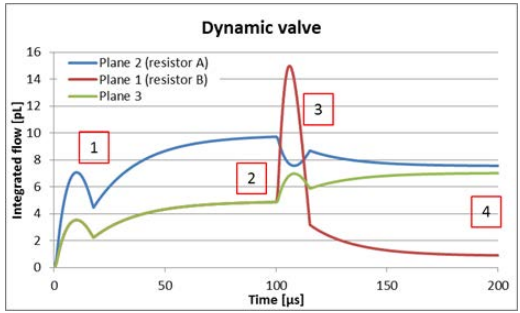

**Figure 14.** *Demonstration of the dynamic valve operation: 1) equal flow propagating to both channels after 10 μs of pump A firing, 2) steady state after 99 μs, 3) flows after 10 μs of pump B firing, 4) steady state at 200 μs.*

The basic strategy is to use pumps to minimize fluid down specific channels of interest. We use pumps and fire them so that they can pump fluid out of channels if previous pumping events have pumped unwanted fluid into these channels. While these valves may be leaky, CFD simulations show that we can greatly minimize flow into specified channels using the same resistors that we used for pumping and mixing.

### *Filters*

Microfilters are important components of microfluidic systems and are a traditional feature used in inkjet applications [3]. SU8 microfabrication technology enables a wide variety of filtering elements, an example is shown in Figure 15.

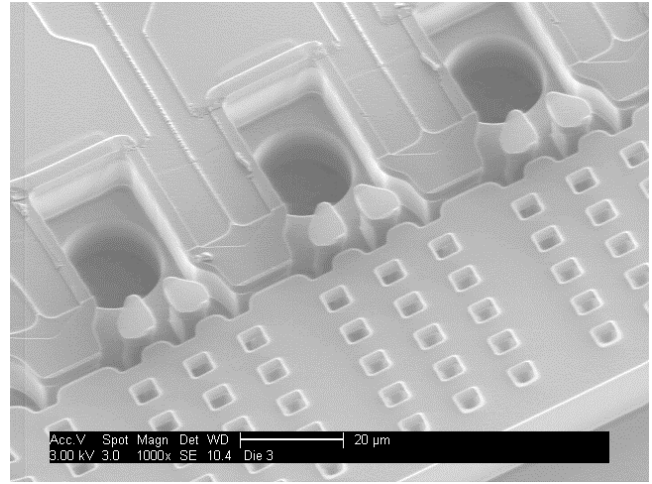

**Figure 15**. *Micrograph of the exemplary SU8 MEMS filters typical for inkjet application protecting ejection device from particle contamination.*

## Conclusion: competitive advantages of inkjet based microfluidic components

New components for integrated microfluidics such as inertial pumps, mixers, flow valves and filters that are described here can be fabricated in the same way and with the same processes as actuators and microchannels in TIJ printhead chips [3]. That implies dozens and hundreds of pumps and other components per chip at a cost of a fraction of a penny per individual component, all without moving parts and individually controlled, if needed, by CMOS circuitry.

Comparison of the bubble-driven inertial micropump with those existing on the market and in research labs can be done using flow rate normalized to pump volume [9]. Using this figure of merit demonstrates significant advantages of the inertial micropump technology and its potential for integration into microfluidic systems due to its high delivered flow rate and small form factor (Figure 16).

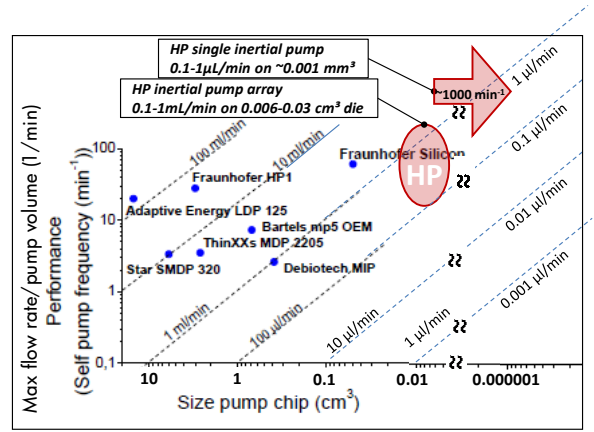

**Figure 16.** *Comparison of the HP inertial micropump with available on the market micro-pumps.*

In addition, the inertial bubble micropump may open avenue for highly integrated and multi-parallel integrated microfluidic networks.

Evaluation of the active micromixers based on TIJ resistor actuators demonstrated $10^2$-$10^4$ mixing efficiency increase with simultaneous $10^2$-$10^3$x size improvement with respect to diffusion driven passive microfluidic mixers. Comparison of performance of the state-of-the-art mixers is shown in Figure 17 upper demonstrating superiority of TIJ based micromixer.

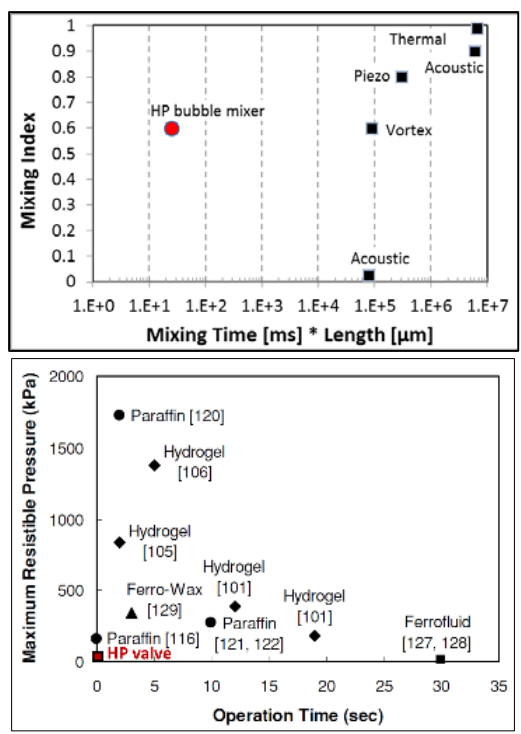

**Figure 17.** *Competitive analysis of micro-mixer (upper) and HP dynamic valve vs other non-Mechanical valves based on [17] (bottom).*

Dynamic microvalve technology comparison with reference to other valve concepts is shown in Figure 17 bottom. The dynamic microvalve has a very fast operating time (on the order of microseconds) and can be placed anywhere in the microfluidic system with no flow losses. Disadvantages for the microvalve is that it is leaky and not able resist high pressure.

Demonstrated step-function improvement in performance of micro-pump, mixer and valve technology based on TIJ platform will certainly usher in a new generation of microfluidic devices. Examples of such devices include polymerase chain reactors, micro-calorimeters, and general pump networks (Figure 18). In the near future, one can envision proliferation of such devices on stationary (printer size) or mobile (cell phone size) platforms, performing various physical, chemical and biological functions.

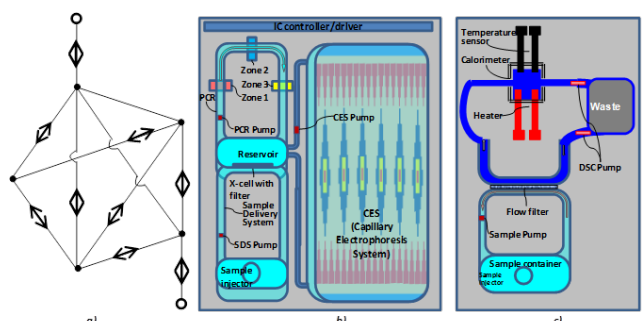

**Figure 18:** *Schematic diagram of: a) fluidic network with reversible pumps, b) lab-on-chip polymerase chain reactor and c) complete micro-calorimeter micro-total analysis system which can be used for protein differential scanning calorimetry or precision DNA melting*

In general, we believe the new integrated microfluidics technology opens up a significant opportunity in customized low cost microfluidic devices. By leveraging highly-developed inkjet printhead assembly processes, we can provide customizable microfluidic devices using pick-and-place technology with either glass or plastic substrates containing resistive/conductive components. All preassembled components can be packaged on solid substrate forming an integrated microfluidic device, which is robust, inexpensive and high-performing.

## Acknowledgements

The authors wish to thank Ken Vandehey for fruitful discussion of the paper and Galen Cook and Paul Richards for SEM images.

## References

[1] G. Whitesides, "The origins and the future of microfluidics," Nature, vol. 442, pp. 368-373, 2006.

[2] "Microfluidic Technologies: Biopharmaceutical and Healthcare Applications 2013-2023," Visiongain, Pharma report, 2013. http://www.visiongain.com/Press_Release/436/Microfluidic-technologies-market-will-reach-3-5bn-in-2017-predicts-new-visiongain-report

[3] J. Stasiak, S. Richards and P. Benning, "Hewlett-Packard's MEMS technology: Thermal inkjet printing," In "Microelectronics to Nanoelectronics: Materials, Devices & Manufacturability" edited by Anupama B. Kaul. CRC Press: pp. 61-78, 2012.

[4] T. Hua, E.D. Torniainen, D.P. Markel and R.N.K. Browning, "Numerical simulation of droplet ejection of thermal inkjet printheads," Int J Numer Meth Fluids, vol.77, pp. 544-570, 2015.

[5] E.D. Torniainen, A.N. Govyadinov, D.P. Markel and P.E. Kornilovitch, "Bubble-driven inertial micropump," Phys Fluids vol. 24, 122003, 2012.

[6] P.E. Kornilovitch, A.N. Govyadinov, D.P. Markel, E.D. Torniainen, "One-dimensional model of inertial pumping," Phys Rev E, vol. 87, 023012, 2013.

[7] A. Nisar , N. Afzulpurkar, B. Mahaisavariya and A. Tuantranont, "MEMS-based micropumps in drug delivery and biomedical applications," Sensors and Actuators B, vol. 130, pp. 917–942, 2008.

[8] M. Nabavi, "Steady and unsteady flow analysis in microdiffusers and micropumps: a critical review," Microfluid Nanodluid, vol. 7, pp. 599-619, 2010.

[9]  D.J. Laser and J.G. Santiago, "A review of micropumps," J. Micromech. Microeng., vol. 14, pp. R35-R64, 2004.

[10] A.K. Au, H. Lai, B.R. Utela and A. Folch, "Microvalves and micropumps for BioMEMS," Micromachines, vol. 2, pp.179-220, 2011.

[11] F.Amirouche, Y. Zhou and T. Johnson, "Current micropump technologies and their biomedical applications," Microsyst. Technol., vol. 15, pp. 647-666, 2009.

[12] W. R. Knight, "Computer modeling of a thermal inkjet device," Proceedings of the IS&T 7th International Congress on Adv. in Non-impact Print. Technol., vol. **7**, pp. 86–95, 1991.

[13] L. Capretto, W. Cheng, M. Hill and X Zhang, "Micromixing within microfluidic devices," Microfluidics, vol. 304, pp. 27-68, 2011.

[14] C.-Y. Lee, C.-L. Chang, Y.-N. Wang and L.-M. Fu, "Microfluidic Mixing: A Review," Int. J. Mol. Sci., vol. 12, pp. 3263-3287, 2011.

[15] D. Markel *et al*. Bubble driven micromixer, to be submitted

[16] J.A. Weaver, J. Melin, D. Stark, S.R. Quake and M.A. Horowitz, "Static control logic for microfluidic devices using pressure-gain valves," Nat. Phys., vol. 6, pp. 218-223, 2010.

[17] K.W. Oh and C.H. Ahn, "A review of microvalves," J. Micromech. Microeng., vol.16,pp. R13–R39, 2006.

## Authors Biography

*Alexander Govyadinov, Ph.D.*, has over 30 years of experience in various sensing platform development in academic and R&D industrial environments, and recent 14 years works for Hewlett-Packard Technology Development Organization. He developed Light Scattering Drop Detection concept implemented in HP first page wide array printers Office jet ProX series. Recently he started working on novel concept of microfluidic architectures enabling advanced inks. He is co-author of multiple scientific publications and over 80 US Patents and patent applications.

*Pavel Kornilovitch* holds a PhD in physics from King's College London. During his 16 years at HP, he worked on a variety of emerging technologies including reflective displays, MEMS sensors, novel materials and nanotechnology. Currently, Pavel is involved in the development of novel microfluidic components. He co-authored over 70 scientific publications and holds over 40 US patents.

*Erik Torniainen* is a senior member of the modeling and simulation group in the Printing and Technology Development Organization at HP (PPS/IPS/PTD/EMAG). He has worked at HP since 1996 and has numerous patents and publications in the microfluidics, heat transfer and computational fluid dynamics space. Erik has a PhD in Mechanical Engineering from Cornell University.

*David P. Markel* has 22 years of experience at Hewlett Packard and works in the Engineering Modeling and Analysis Group (EMAG). He is the curator of HPs proprietary computational fluid dynamics application along with providing support to a variety of IPS programs. David holds 15 patents. A lifelong learner, David received a bachelor's degree in information systems from Linfield College and also attained degrees in electronics engineering, water & environmental technology, and physics.